\documentclass[twocolumn,preprintnumbers,amsmath,amssymb]{revtex4}

\usepackage{graphicx}
\usepackage{dcolumn}
\usepackage{bm}

\newcommand{\beq}{\begin{equation}}
\newcommand{\eeq}{\end{equation}}
\newcommand{\bea}{\begin{eqnarray}}
\newcommand{\eea}{\end{eqnarray}}
\newcommand{\bdm}{\begin{displaymath}}
\newcommand{\edm}{\end{displaymath}}

\def\as{\alpha_s}

\def\m{{\cal M}}

\def\ord{{\cal O}}

\def\dy{\Delta y}

\begin{document}

\preprint{MAN/HEP/2009/21}

\title{Gaps between jets and soft gluon resummation\footnote{Talk given at ``London workshop on Standard Model Discoveries with early LHC data'', UCL, London, 31$^{\rm st}$ March-1$^{\rm st}$ April 2009. \\Also given at ``DIS 2009'', Madrid, Spain, 26$^{\rm th}$-30$^{\rm th}$ Apr 2009 }}

\author{Simone Marzani}
 \email{simone.marzani@manchester.ac.uk}
 
 \author{Jeffrey Forshaw}
 \email{jeff.forshaw@manchester.ac.uk}

 \author{James Keates}
 \email{james.keates@cern.ch}

 \affiliation{School of Physics \& Astronomy, University of Manchester,\\Oxford Road, Manchester, M13 9PL, U.K.}




\begin{abstract}
We study the effect of soft gluon resummation on the gaps-between-jets cross-section at the LHC. We review the theoretical framework that enables one to sum logarithms of the hard scale over the veto scale to all orders in perturbation theory. We then present a study of the phenomenological impact of Coulomb gluon contributions and super-leading logarithms on the gaps between jets cross-section at the LHC.
\end{abstract}

\maketitle

\section{Jet vetoing: gaps between jets}
We consider dijet production with transverse momentum $Q$ and a veto on the emission of additional radiation in the inter-jet rapidity region, $Y$, harder than $Q_0$.
 We shall refer generically to the ``gaps between jets'' process, although the veto scale is chosen to be large, $Q_0= 20$~GeV, so that we can rely on perturbation theory. Thus  a ``gap'' is simply a region of limited hadronic activity. 

Gaps between jets is a pure QCD process, hence the cross-section is large and studies can be performed with early LHC data. It is interesting because it allows one to investigate a remarkably diverse range of QCD phenomena. 
For instance, the limit of large rapidity separation corresponds to the limit of high partonic centre of mass energy and  BFKL effects are expected to become important~\cite{muellernavelet}.  On the other hand one can study the limit of emptier gaps, becoming more sensitive to wide-angle soft gluon radiation. Furthermore, if one wants to investigate both of these limits simultaneously, then  the non-forward BFKL equation enters the game~\cite{muellertang}.  In the following we discuss only wide-angle soft emissions.

Accurate studies of these effects are important also in relation to other processes, in particular the production of a Higgs boson in association with two jets. It is well known that this process can occur via gluon-gluon fusion and weak-boson fusion (WBF). QCD radiation in the inter-jet region is clearly  different in the two cases and, in order to enhance the WBF channel, one can put a cut on emission between the jets \cite{Barger:1994zq,Kauer:2000hi}. This situation is very closely related to gaps between jets since the Higgs carries no colour charge, and QCD soft logarithms can be resummed using the same technique~\cite{softgluonshiggs}. 

Given a hard scattering process, we can study how it is modified by the addition of soft radiation. If the observable is inclusive enough, then we  have no effects because soft contributions cancel when real and virtual corrections are added together, as a result of the Bloch-Nordsieck theorem. However, if we restrict the real radiation to a corner of the phase space, as happens for the gap cross-section, we encounter a miscancellation and are left with a logarithm of the ratio of the hard scale and veto scale, $Q/Q_0$.
The resummation of wide-angle soft radiation in the gaps between jets process was originally performed assuming that the real--virtual cancellation is perfect outside the gap, so that one needs only to consider virtual gluon corrections integrated over momenta for which real emissions are forbidden, i.e. over the ``in gap'' region of rapidity and with
$k_T$ above the veto scale $Q_0$~\cite{KOS,OS, Oderda}. We shall refer to these contributions as global logarithms. The resummed squared matrix element can be written as:
\bea \label{resummedpartonicxsec1}
|\m|^2 &=& \frac{1}{V_c} \langle m_0 | e^{- \xi \mathbf{\Gamma}^{\dagger}}e^{- \xi \mathbf{\Gamma}} |m_0 \rangle\,, \nonumber \\
 \xi  &=&\frac{2}{\pi} \int_{Q_0}^{Q} \frac{d k_T}{k_T} \as(k_T)\,,
\eea
where $V_c$ is an averaging factor for initial state colour.
The vector $|m_0 \rangle$ represents the Born amplitude and the operator $\mathbf{\Gamma}$ is the soft anomalous dimension:
\beq \label{gammaoperator}
\mathbf{\Gamma} = \frac{1}{2}Y \mathbf{t}_t^2+ i \pi \mathbf{t}_a\cdot \mathbf{t}_b +\frac{1}{4}\rho_{\rm jet}(Y, |\dy|)(\mathbf{t}_c^2+\mathbf{t}_d^2)\,,
\eeq
where $\mathbf{t}_i$ is the colour charge of parton $i$ and the function $\rho_{\rm jet}(Y,\dy)$ is related to the jet definition.
The operator $\mathbf{t}_t^2$ represents the colour exchanged in the $t$-channel:
\beq
\mathbf{t}_t^2=(\mathbf{t}_a+\mathbf{t}_c )^2 = \mathbf{t}_a^2+\mathbf{t}_c^2 + 2  \,\mathbf{t}_a\cdot \mathbf{t}_c\,.
\eeq
The imaginary part of Eq.~(\ref{gammaoperator}) is due to Coulomb gluon exchange. These contributions play an important role in the proof of QCD factorization and they are also responsible for super-leading logarithms~\cite{SLL1,SLLind}. We notice that for processes with less than four coloured particles, such as deep-inelastic scattering or Drell-Yan processes, the imaginary part of the anomalous dimension does not contribute to the cross-section. For instance, if we consider three coloured particles, then colour conservation implies that $ \mathbf{t}_a+\mathbf{t}_b + \mathbf{t}_c=0$, and consequently 
\beq
 i\pi \, \mathbf{t}_a \cdot \mathbf{t}_b =\frac{i\pi}{2} \left( \mathbf{t}_c^2 -\mathbf{t}_a^2-\mathbf{t}_b^2 \right)\,,
 \eeq
which contributes as a pure phase. Coulomb gluons do play a role in dijet production, but they are not implemented in angular-ordered parton showers. We shall evaluate the impact of these contributions on the cross-section in the next section.

It was later realised~\cite{DS} that the above procedure is not enough to capture the full leading logarithmic behaviour. Real gluons emitted outside of the gap are forbidden to re-emit back into the gap and this gives rise to a new tower of logarithms, formally as important as the primary emission corrections, known now as non-global logarithms.
 The leading logarithmic accuracy is therefore achieved by considering all $2 \to n $ processes, i.e. \mbox{$n-2$ out-of-gap gluons}, dressed with ``in-gap'' virtual corrections, and not only the virtual corrections to the $2\to 2$ scattering amplitudes.
The colour structure quickly becomes intractable and, to date, calculations have been performed only in the large $N_c$ limit~\cite{DS,appleby2,nonlinear}.

A different approach was taken in~\cite{SLL1,SLLind}, where the specific case of only one gluon emitted outside the gap, dressed to all orders with virtual gluons but keeping the full $N_c$ structure, was considered. That calculation had a very surprising outcome, namely the discovery of a new class of ``super-leading'' logarithms (SLL), formally more important than the ``leading'' single logarithms.
Their origin can be traced to a failure of the DGLAP ``plus-prescription'', when the out-of-gap gluon becomes collinear to one of the incoming partons. Real and virtual contributions do not cancel as one would expect and one is left with an extra logarithm.  This miscancellation  first appears at the fourth order relative to the Born cross-section and it is caused by the imaginary part of loop integrals, induced by Coulomb gluons. These SLL contributions  have been recently resummed to all orders in~\cite{jetvetoing}. The result takes the form:
\beq \label{master2}
|\m_1^{\rm SLL}|^2 =   - \frac{2 }{\pi} \int_{Q_0}^{Q} \frac{d k_T}{k_T}\as(k_T) \left(  \ln \frac{Q}{k_T} \right)  \left( \Omega^{\rm coll}_R + \Omega^{\rm coll}_V\right), 
\eeq  
where $ \Omega^{\rm coll}_{R(V)}$ is the resummed real (virtual) contribution in the limit where the out-of-gap gluon becomes collinear to one of the incoming partons. The presence of SLL has been also confirmed by a fixed order calculation in~\cite{SLLfixed};  in this approach SLL have been computed at $\ord (\as^5)$ relative to Born, i.e. going beyond the one out-of-gap gluon approximation.
 
\section{LHC phenomenology}
In this section we perform two different studies. Firstly we consider the resummation of global logarithms and we study the importance of Coulomb gluon contributions, comparing the resummed results to the ones obtained with a parton shower approach. We then turn our attention to SLL and we evaluate their phenomenological relevance. In both studies we consider $\sqrt{S}=14$~TeV, $Q_0=20$~GeV, jet radius $R=0.4$ and we use the MSTW 2008 LO parton distributions~\cite{mstw08}. 
\begin{figure*}
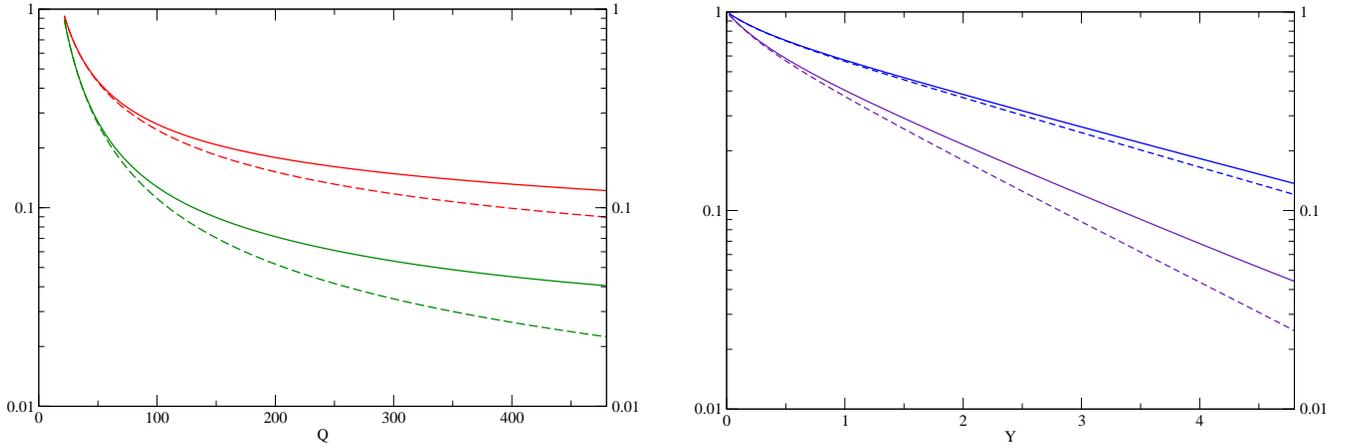

\begin{center}
\includegraphics[width=0.47\textwidth]{marzani_simoneGAP.fig1.eps} \hspace{.5cm}
\includegraphics[width=0.47\textwidth]{marzani_simoneGAP.fig2.eps}
\caption{On the left we plot the gap fraction for $Y=3$ (upper red curves) and $Y=5$ (lower green curves) as a function of $Q$ and on the right as a function of $Y$, for $Q=100$~GeV (upper blue curves) and $Q=500$~GeV (lower violet curves). The solid lines are the full resummation of global logarithms, while the dashed ones are obtained by omitting the $i \pi$ terms in the anomalous dimension.}\label{fig:kfact1}
\end{center}
\end{figure*}

\begin{figure*}
\begin{center}
\includegraphics[width=0.65\textwidth]{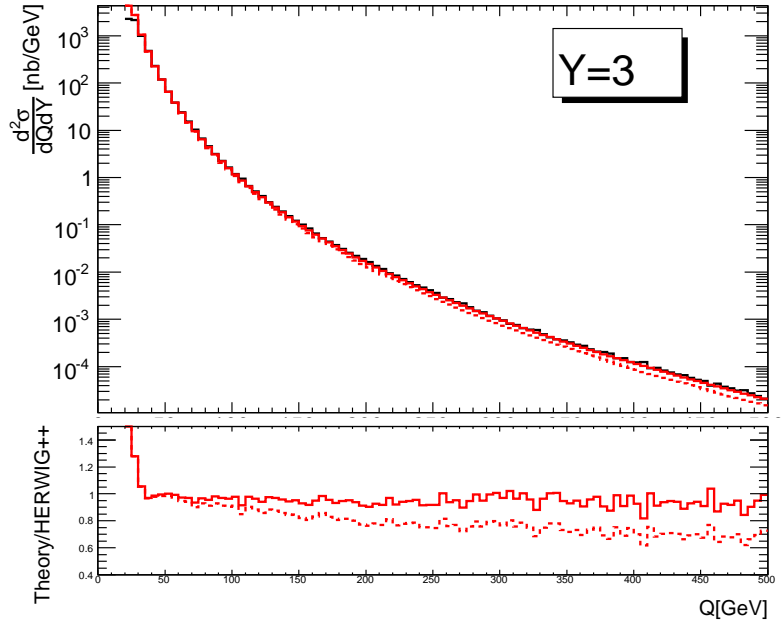}
\caption{The gap cross-section obtained using \textsc{Herwig++} (black histogram) is compared to the one from resummation (red curves). As before the solid line is the full result, while the dashed line is obtained by omitting the Coulomb gluon contributions. 
At the bottom we plot the ratio between the results obtained from the resummation and the one from \textsc{Herwig++}.}\label{fig:gapx}
\end{center}
\end{figure*}

Soft logarithmic contributions are implemented in \textsc{Herwig++} via angular ordering of successive emissions. Such an approach cannot capture the contributions coming from the imaginary part of the loop integrals, due to Coulomb gluon exchange. We evaluate the importance of these contributions in Fig.~\ref{fig:kfact1}. On the left  we plot the gap cross-section, normalised to the Born cross-section (i.e. the gap fraction), as a function of $Q$ at two different values of $Y$ and, on the right, as a function of $Y$ at two different values of $Q$. The solid lines represent the results of the resummation of global logarithms; the dashed lines are obtained by omitting the $i \pi$ terms in the soft anomalous dimension matrices.  As a consequence, the gap fraction is reduced by $7\%$ at $Q=100$~GeV and $Y=3$ and by as much as $50\%$ at $Q=500$~GeV and $Y=5$. Large corrections from this source herald the breakdown of the parton shower approach.  In Fig.~\ref{fig:gapx} we compare the gap cross-section obtained after resummation  to that obtained using \textsc{Herwig++}~\cite{ThePEG,Bahr:2008pv,KLEISSCERN9808v3pp129,Gieseke:2003rz} after parton showering ($Q$ is taken to be the mean $p_T$ of the two leading jets). The broad agreement is encouraging and indicates that effects such as energy conservation, which is included in the Monte Carlo, are not too disruptive to the resummed calculation. Nevertheless, the histogram ought to be compared to the dotted curve rather than the solid one, because \textsc{Herwig++} does not include the Coulomb gluon contributions. 
The resummation approach and the parton shower differ in several aspects: some non-global logarithms are included in the Monte Carlo and the shower is performed in the large $N_c$ limit. Of course the resummation would benefit from matching to the NLO calculation and this should be done before comparing to data.  

\begin{figure*}
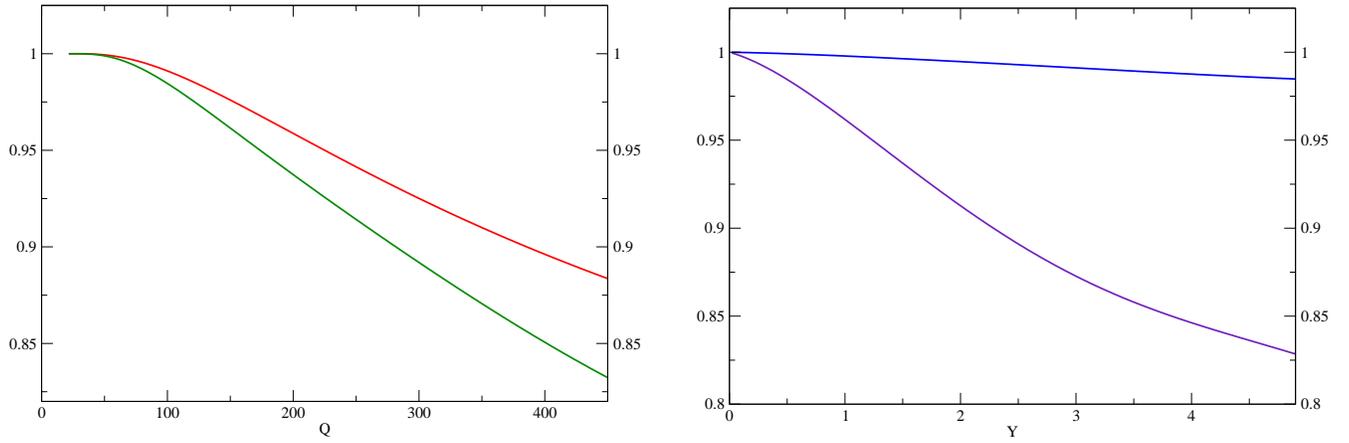

\begin{center}
\includegraphics[width=0.47\textwidth]{marzani_simoneGAP.fig4.eps} \hspace{.5cm}
\includegraphics[width=0.47\textwidth]{marzani_simoneGAP.fig5.eps}
\caption{On the left we plot the $K$-factor as defined in Eq.~(\ref{kSLL}) as a function of $Q$ for $Y=3$ (upper red curve) and $Y=5$ (lower green curve);  on the right we plot it as a function of $Y$, for $Q=100$~GeV (upper blue curve) and $Q=500$~GeV (lower violet curve) }\label{fig:kSLL}
\end{center}
\end{figure*}

Finally we want to study the relevance of the SLL contributions. In order to do that we define
\beq \label{kSLL}
K^{(1)}= \frac{\sigma^{(0)}+\sigma^{(1)}}{\sigma^{(0)}} \,,
\eeq
where $\sigma^{(0)}$ contains the resummed global logarithms and  $\sigma^{(1)}$ the resummed SLL contribution coming from the case where one gluon is emitted outside of the gap.
The results are shown in Fig.~\ref{fig:kSLL}.
Generally the effects of the SLL are modest, reaching as much as 15\% only for jets with \mbox{$Q > 500$~GeV} and rapidity separations $Y > 5$. The contribution coming from $n \ge 2$ out-of-gap gluons is thought to be less important~\cite{jetvetoing}.
Remember that we have fixed the value of the veto scale $Q_0=20$~GeV and that the impact will be more pronounced if the veto scale is lowered. 
\section{Conclusions and Outlook}
There is plenty of interesting QCD physics in ``gaps-between-jets'' and measurement can be performed with early LHC data.  
There are significant contributions arising from the exchange of Coulomb gluons, especially at large $Q/Q_0$ and/or large $Y$, which are not implemented in the parton shower Monte Carlos. However, before comparing to data, there is a need to improve the resummed results by matching to the fixed order calculation. These observations will have an impact on jet vetoing in Higgs-plus-two-jet studies at the LHC.

We have studied the super-leading logarithms that occur because gluon emissions that are collinear to one of the incoming hard partons are forbidden from radiating back into the veto region. Even if their phenomenological relevance is generally modest, they deserve further study because they are deeply connected to the fundamental ideas behind QCD factorization.
 
\newpage

\begin{footnotesize}

\end{footnotesize}


\end{document}